\documentclass[prl,twocolumn,showpacs,preprintnumbers,amsmath,amssymb,unsortedaddress]{revtex4-1}

\usepackage{dcolumn}
\usepackage{bm}

\usepackage{subfigure}
\usepackage[pdftex]{graphicx}

\usepackage{ulem}
\usepackage{color}

\begin{document}

\title{Universal flow-density relation of single-file bicycle,
pedestrian and car motion}

\author{J. Zhang}
\author{W. Mehner}
\author{S. Holl}
\author{M. Boltes}
\affiliation{J\"ulich Supercomputing Centre, Forschungszentrum
  J\"ulich GmbH, 52425 J\"ulich, Germany}
\email{ju.zhang@fz-juelich.de}
\email{w.mehner@fz-juelich.de}
\email{st.holl@fz-juelich.de}
\email{m.boltes@fz-juelich.de}


\author{E. Andresen}
\affiliation{Department of Computer Simulation for Fire Safety and Pedestrian Traffic,
  Bergische Universit\"at Wuppertal, 42285 Wuppertal, Germany}
\email{e.andresen@uni-wuppertal.de}



\author{A. Schadschneider}
\affiliation{Institut f\"ur Theoretische Physik, Universit\"at zu
  K\"oln, 50937 K\"oln, Germany}
\email{as@thp.uni-koeln.de}

\author{A. Seyfried}
\affiliation{J\"ulich Supercomputing Centre, Forschungszentrum
  J\"ulich GmbH, 52425 J\"ulich, Germany}
\affiliation{Department of Computer Simulation for Fire Safety and Pedestrian Traffic,
Bergische Universit\"at Wuppertal, 42285 Wuppertal, Germany}
\email{a.seyfried@fz-juelich.de}

 \begin{abstract}

   The relation between flow and density is an essential quantitative characteristic to
   describe the efficiency of traffic systems.  We have performed
   experiments with single-file motion of bicycles and compare the
   results with previous studies for car and pedestrian motion in
   similar setups. In the space-time diagrams we observe three
   different states of motion (free flow state, jammed state and
   stop-and-go waves) in all these systems.  Despite of their
   obvious differences they are described by a universal
   fundamental diagram after proper rescaling of space and time which
   takes into account the size and free velocity of the three kinds of
   agents. This indicates that the similarities between the systems go
   deeper than expected.

\end{abstract}



\maketitle

\section{Introduction}

In the past, various studies have been performed on pedestrian
\cite{Schadschneider2009a}, bicycle \cite{Taylor1999} and vehicular
traffic \cite{Chowdhury2000,Nagatani2002a,Kerner2004,Treiber2013}. Besides
the obvious practical relevance, from a physics point-of-view, these
traffic systems are interesting for the observed collective and
self-organization phenomena, phase transitions etc. Most of the
methods and theories in pedestrian dynamics are borrowed from
vehicular traffic.  As for the study of bicycle traffic, most research
focuses on operating characteristics, travel speed distributions as
well as bicycle characteristics. Only a small number of studies
focused on the flow properties of bicycle traffic
\cite{Smith1976,Navin1994,Jiang2011a,Andresen2013}.
Here we want to find out how strongly the flow-density relation
depends on the properties of the agents.

Usually these different types of traffic flows are investigated
separately. So far a systematic comparison has not been attempted but
qualitative similarities are obvious. Nearly all studies on
pedestrians and vehicles show that the speed decreases with the
density. At a certain critical value of the density the flow is
unstable and transits from free flow to jammed flow. This transition
was also found for bicycle flows \cite{Andresen2013}. In this work
single-file pedestrian, car and bicycle movement on a planar
circuit will be studied under laboratory conditions. We analyze, on
a quantitative level, similarities and differences between the
flow-density relation of these three traffic modes. We want to study
whether they can be derived from a
universal flow-density relation.


\section{Experimental Setup}

The experiments for all three types of traffic were performed with
similar setups, namely on planar circuits where only single-file
motion was possible. Series of experiments were carried out with a maximal number of
participants $N = $ 70, 23 and 33 for the pedestrian, car and bicycle
experiment, respectively. In general, participants were asked to
move normally without overtaking.  The global density was varied by
repeating the experiment with different numbers of participants.

The pedestrian experiment \cite{Seyfried2010b} was performed with
soldiers moving in a circular corridor of circumference $C_p = 26~$m.
During the experiment the soldiers were asked to walk in a normal fashion but not in lockstep (see the video from \cite{pedvideo}).The one dimensional global density $\rho_g = N/C$ ranges from 0.54~m$^{-1}$ to
2.69~m$^{-1}$ in this experiment.

A similar experiment with cars was performed by Sugiyama et al.
\cite{Sugiyama2008,Nakayama2009} on a circular road with circumference
$C_c=230~$m and $N = 22$ and 23, corresponding to global densities
$\rho_g=0.096$~m$^{-1}$ and $0.1$~m$^{-1}$, respectively.
Recently the same group improved these experiments \cite{Tadaka2013}. They carried out 19
experimental runs with $C_c=312~$m and different numbers of cars
($N$ was changed from 10 to 40). The global density $\rho_g$ in this
experiment ranges from 0.03~m$^{-1}$ to 0.13~m$^{-1}$.

\begin{figure}
\centering{
\includegraphics[width=0.9\columnwidth]{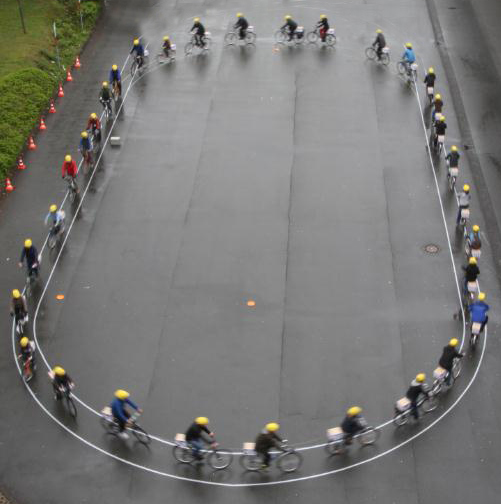}}
\caption{\label{fig-car-bike-exp} Snapshots of the bicycle
experiment on a circuit road.}
\end{figure}

The bicycle experiment was carried out in Germany in 2012 with
participants of all ages \cite{Zhang2013}. On a circuit road with
circumference $C_b=86~$m several runs with different numbers of
bicycles (from $N = 5$ to $N=33$) were performed
(Fig.~\ref{fig-car-bike-exp}).  Based on video recordings \cite{bikevideo} the
trajectories were extracted automatically, similar to the method used
in the pedestrian experiments \cite{Boltes2010}.  Details will be
given elsewhere \cite{Zhang2013}.

\section{Trajectories}

\begin{figure}
\centering\subfigure{
\includegraphics[width=0.95\columnwidth]{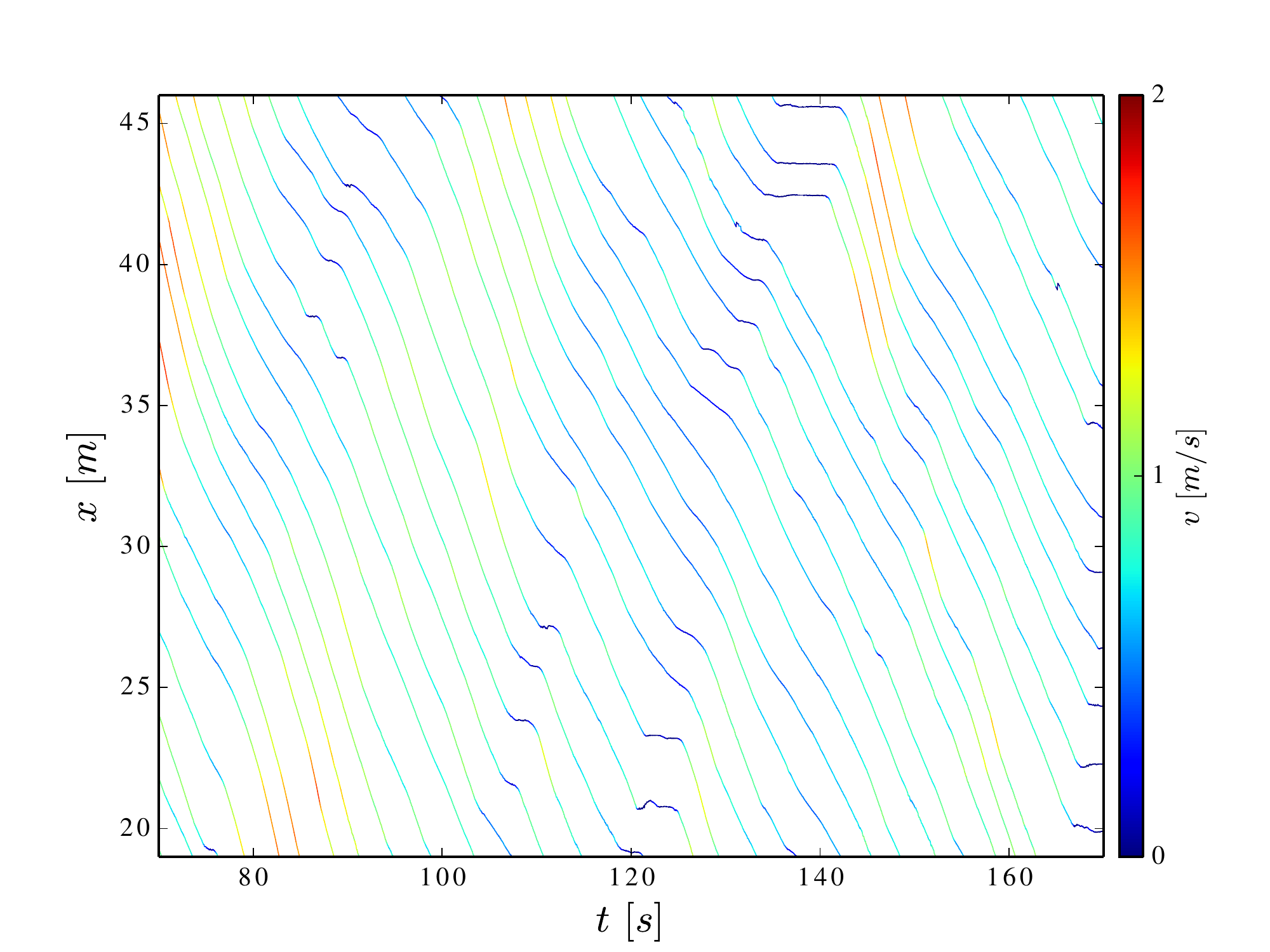}}
\caption{\label{fig-timespace-bike} Trajectories in the measurement
  area (of length 27~m) for the bicycle experiment with $N=33$. The same structures can be found in trajectories of pedestrian and vehicle systems \cite{Sugiyama2008,Seyfried2010b}.}
\end{figure}

From the high precision trajectories, traffic flow characteristics
including flow, density and velocity can be determined.
Fig.~\ref{fig-timespace-bike} shows a time-space diagram in the
measurement area, which has a length of $27~$m, from one run of the
bicycle experiment.  Similar trajectory plots for car and pedestrian
motion can be found in \cite{Sugiyama2008,Seyfried2010b}.

In all three cases a transition from free flow to jammed flow can be
observed with the increasing of the global density. In the free flow
regime all agents can move at their desired speed, whereas in the
jammed regime typically stop-and-go waves are observed.

\section{Methods of data analysis}

The comparison of time-space diagrams already indicates a qualitative
similarity between the three traffic systems.  A deeper understanding
requires quantitative analysis which allows to uncover the underlying
dynamics that is not apparent in the more qualitative observations
of the trajectories.  We use both macroscopic and microscopic analysis
to obtain more detailed information on the
specific flow-density relation or equivalently the velocity-density relation.
In this study  the specific flow $J_s$ is calculated by $J_s = \rho \cdot v$.

Microscopically, an individual density can either be defined based on a
Voronoi tesselation \cite{Steffen2010a} or the headway. The headway $d_H(i)$
is  defined as the distance between the centers of mass of an agent $i$ and
its predecessor, whereas the Voronoi space $d_V(i)$, for one-dimensional
motion, is the distance between the midpoints of the headway and the headway
of its follower $i-1$.  The corresponding individual densities are then
 $\rho_H(i) = 1/d_H(i)$ and $\rho_V(i)= 1/d_V(i)$, respectively. As for the
 individual velocity, the instantaneous velocity $v_i(t)$ is defined as
\begin{equation}\label{eq2}
v_i(t)=\frac{{x_i}(t+\Delta t^\prime/2)-{x_i}(t-\Delta
t^\prime/2))}{\Delta t^\prime}\,,
\end{equation}
where $x_i(t)$ is the x coordinate of pedestrian $i$ at time $t$ and
$\Delta t^\prime = 2~$s is used in this study.

The individual flow-density relation of bicycle traffic obtained from
the Voronoi-based and headway-based methods do not show large
discrepancies but the results of the headway-based method are more
scattered. This has previously been
observed for pedestrian dynamics \cite{Steffen2010a,zhang2011}.

Macroscopically, the similarities between the three systems are more apparent.
On the macroscopic level the mean densities $\rho (t)$ and velocities
$v (t)$ in a measurement area at time $t$ are calculated based on Voronoi
method:

\begin{eqnarray}\label{eq1}
\rho (t)&=&\frac{\sum_{i=1}^{n}\Theta_i(t)}{l_m}\,,\\
v(t)&=&\frac{\sum_{i=1}^{n}{\Theta_i(t)\cdot v_i(t)}}{l_m}\,,
\end{eqnarray}
where $n$ is the number of agents whose Voronoi space includes the
measurement area (assuming that the overlapping length between the space and the measurement area is $d_{o}(i)$ for
agent $i$).  $\Theta_i(t) = d_{o}(i)/ d_{V}(i)$ represents the
contribution of agent $i$ to the density of the measurement area.
$v_i(t)$ is the instantaneous velocity (see eq.~(\ref{eq2})) and $l_m$ is the
length of the measurement area.

In this study, the lengths of the measurement areas $l_m$ were 3~m, 35~m and
13~m in the pedestrian, car and bicycle experiments, respectively. Since the (average) length of agents
is $0.4~$m, $3.9~$m  and $1.73~$m, at most 7 agents can occupy
the measurement area at the same time.
It should be noted that in all experiments the ratio of agent
length and system length was of the same order of magnitude.


\section{Results}

\begin{figure}
\centering{
\includegraphics[width=0.9\columnwidth]{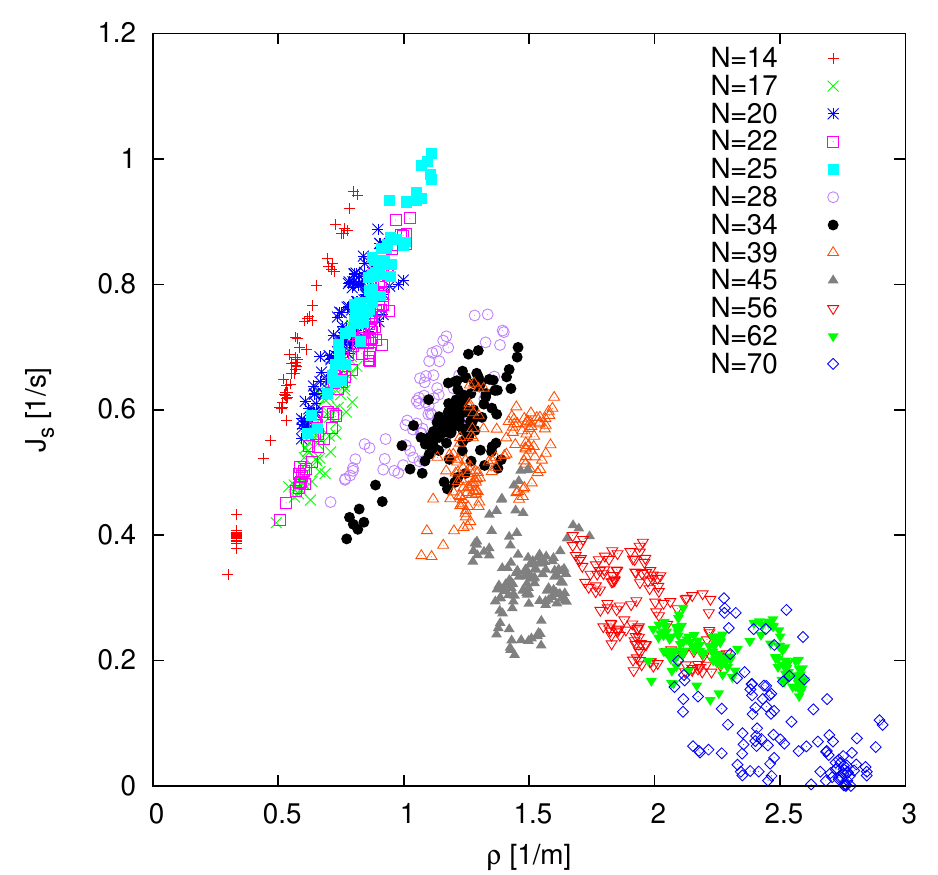}}
\caption{\label{fig-FD-Voro-peds} The Voronoi-based flow-density relation for pedestrians. For $\rho > 1.0~$m$^{-1}$, the free flow
  regime ends and the specific flow decreases with increasing
  density. }
\end{figure}

The flow-density relation of the pedestrian experiment
can be divided into three regimes $\rho \in [0,
1.0]$~m$^{-1}$, $[1.0,1.7]$~m$^{-1}$ and $[1.7, 3.0]$~m$^{-1}$, which
correspond to three states of pedestrian movement (see
Fig.~\ref{fig-FD-Voro-peds}). For small densities
$\rho < 1.0~$m$^{-1}$ free flow is observed and the specific flow
increases monotonically with the density. A gap can be observed in the
velocity-density relationship at $\rho = 1.0~$m$^{-1}$, where a
transition occurs in the specific flow-density relationship. When the number of
pedestrians inside the system increases from 25 to 28, the specific
flow jumps from $1.0$~s$^{-1}$ to $0.7~$s$^{-1}$. For $\rho >
1.0~$m$^{-1}$, the stream is in a congested state and the specific
flow starts to decrease with the increasing density. However, the
decline rates are different around $1.7~$m$^{-1}$. The transient
decelerations of pedestrians can be observed sometimes but are not the
main property of the movement for $\rho < 1.7~$m$^{-1}$.  For $\rho >
1.7~$m$^{-1}$, stop-and-go waves dominate the motion of the
pedestrians, which can be seen in \cite{Seyfried2010b}.

\begin{figure}
  \centering{
    \includegraphics[width=0.9\columnwidth]{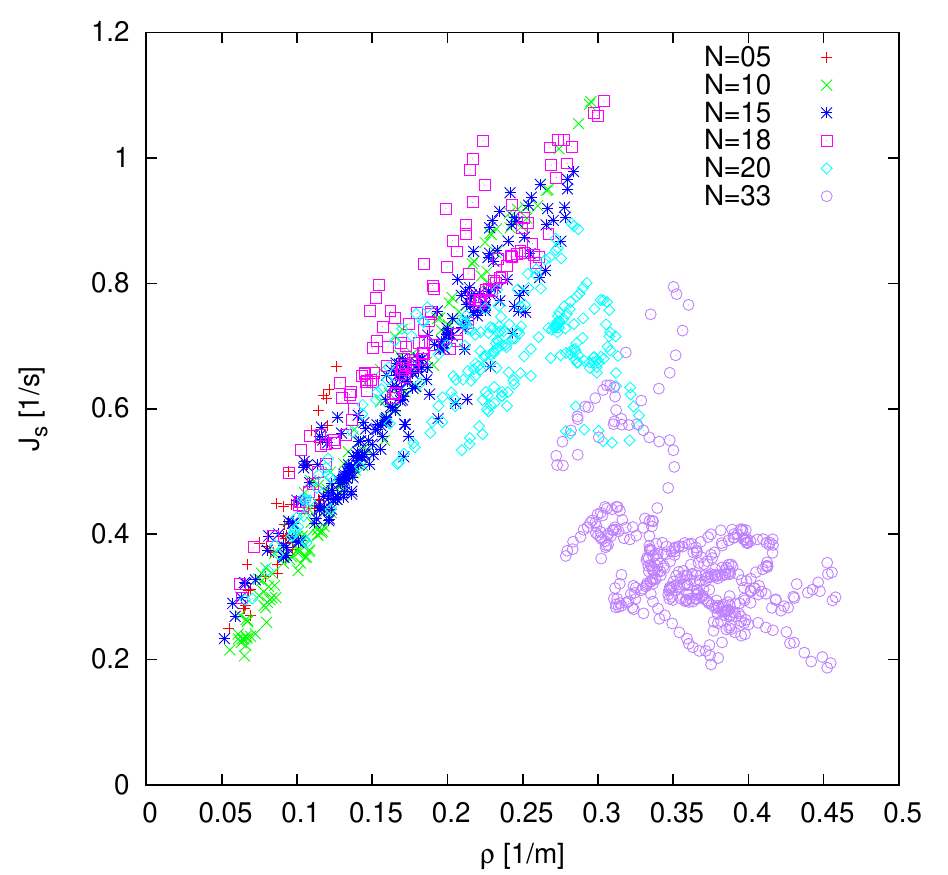}}
\caption{\label{fig-FD-Voro-bike} The Voronoi-based flow-density relation of the bicycle experiment. For $\rho > 0.28~$m$^{-1}$
the free flow regime ends and the flow decreases.}
\end{figure}

Similar results are observed in the bicycle system, as shown in
Fig.~\ref{fig-FD-Voro-bike}. For densities $\rho < 0.3~$m$^{-1}$, the
bicycle stream is in a free flow state and the specific flow increases
monotonically with increasing density. At $\rho = 0.3~$m$^{-1}$ the
specific flow drops sharply from 1.0~s$^{-1}$ to $0.4$~s$^{-1}$
marking the transition to the congested state. Stop-and-go waves
are observed in the run with $N = 33$.


\section{Comparison of the flow-density relations}

Plotting the flow-density relation of these three systems in one graph
shows that the data points occupy different ranges of density and do
not seem to be comparable to each other. To take into account the
different scales of sizes and velocities of the agents we rescale
these quantities. For the length of the agents
we use $L_0(c) = 0.4~$m for pedestrians, $L_0(c) = 3.9~$m for cars \cite{Tadaka2013}
and the mean value of $L_0(b) = 1.73~$m of bicycles in the experiment.
For scaling the speed we consider the desired velocity of each agent.
From special measurements in the course of the experiments we know
that they are about $1.4~$m/s for pedestrians and $5.5~$m/s for
bicycles. For cars here we use $11.1~$m/s (about $40~$km/h) according to the experiment in \cite{Tadaka2013}.

\begin{figure}
\centering{
\includegraphics[width=0.95\columnwidth]{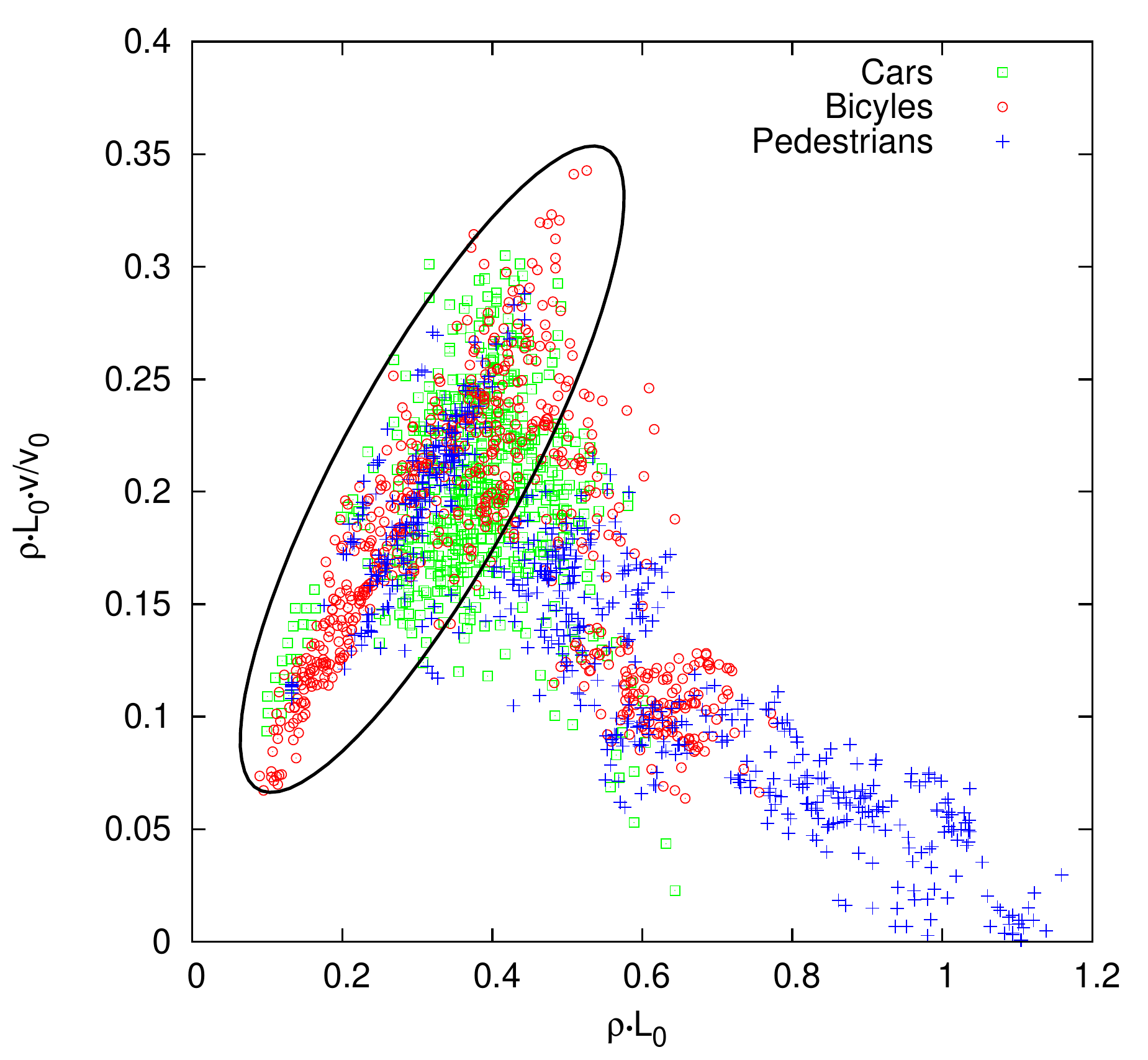}}
\caption{\label{fig-comparison} Comparison of the scaled flow-density relations for pedestrian, car and bicycle traffic.
  For the scaling the desired velocities (1.4, 11.1 and 5.5 m/s) and agent lengths (0.4, 3.9 and  1.73 m)
  have been used respectively. The ellipse shows the outline of the free flow regime.}
\end{figure}

After rescaling it is found that the flow-density relations agree well (see Fig.~\ref{fig-comparison}).
In all cases the free flow regimes ends at approximately $\rho\cdot
L_0 =0.5$.  This implies that the transition to the
congested state occurs when nearly 50\% of the available space is
occupied. Moreover, the capacity, i.e. the maximal
flow, agrees for the three sydiagram are
again similar for all three systems. It consists of different states
like synchronized traffic and stop-and-go waves.
For pedestrians and bicycles, the latter occur
at an occupation of 0.7.

Note that the maximum of the scaled densities $\rho \cdot
L_0$ is larger than 1.0 for pedestrians. In these
experiments no notable body contact was observable and the
compressibility of human bodies is not responsible for this effect.
Instead pedestrian trajectories are extracted by detecting markers on
the head. Head movement in combination with evasion to the side and the
projection of the trajectories in one dimension for this comparative
analysis is responsible for values of the rescaled density higher than
1. For bicycles, there could be some overlapping of bikes in a
zipper-like manner at higher densities which leads to densities larger
than 1.  In contrast, such overlapping is impossible for car traffic
since cars have to keep certain distance to avoid potential
collisions. This explains the lack of data for vehicular traffic close
to density 1.

\section{Discussion}

For single-file systems we showed that the flow-density relations of
pedestrian, car and bicycle traffic show a good agreement after
simple rescaling. The density at maximal flow and the corresponding
flow values are almost identical.
This implies that the shape of the flow-density relation does not
contain much information about the type of traffic. This is somewhat
unexpected since the traffic systems appear to be governed by rather
different aspects. Vehicular traffic is dominated by the physical
restrictions on car motion, e.g. inertia effects limiting the possible
accelerations. In contrast, in pedestrian motion acceleration and
deceleration (and even changes in the direction motion) are almost
instantaneous. Bicycle traffic takes an intermediate position between
these two extremes.

The transport properties in such systems could be approximated by the
universal equation $\tilde{v}=1-\tilde{\rho}$ with $\tilde{v}=v/v_0$
and $\tilde{\rho}=\rho/\rho_0$. This leads to a normalized maximal
flow of 0.25 at a relative density of 0.5. This corresponds to the
properties to the asymmetric simple exclusion process \cite{Derrida98,SCNBook}, which is
often considered as a minimal model for traffic flows. The main feature of
this model is volume exclusion. Also models for pedestrian dynamics
\cite{Kirchner2004,Seyfried2005,Chraibi2010} show that these transport
characteristics could be reproduced by an appropriate consideration of
a velocity dependent volume exclusion, which seems a universal
characteristics of such systems.  From this we conclude that other
properties of the agent, like acceleration or inertia are less
relevant for the structure of the flow-density relation in single-file
traffic systems of different agent types. In other words models
without a proper consideration of the volume exclusion miss an
important aspect of traffic systems.

Two exceptions from these observations deserve to be mentioned.
First of all, in highway traffic two different congested phases
can be distinguished, the wide jam phase and the synchronized phase
\cite{Kerner2004}. The structure of synchronized traffic leads to
a non-functional form of the flow-density relation with a non-unique
flow-density relation.

The second exception, which is more relevant for the present study,
concerns traffic on ant trails. As shown in the empirical study
\cite{JohnSCN09} no congested phase exists up to the largest observed
densities $\rho\approx 0.8$. The average velocity is almost
independent of density and the flow-density diagram consists only
of a monotonically increasing free-flow branch.

It is not immediately clear what the origin of this different
behavior is. As found in \cite{JohnSCN09} ants move in
platoons with small headway, but almost identical velocities.

Summarizing, we have shown that the transport properties of these
three different types of single-file traffic flows can be unified in a
certain range by a simple scaling of velocity and density. These results
may not only provide insights into dynamical behavior but also may be
relevant for the improvement of mixed traffic systems. However, to
investigate this point further empirical data is still needed
especially in the higher density range for bicycle traffic and lower
density range for cars.



\end{document}